\documentclass[11pt]{article}
\usepackage{moriond,epsfig}

\def\be{\begin{equation}}
\def\ee{\end{equation}}
\def\bea{\begin{eqnarray}}
\def\eea{\end{eqnarray}}

\begin{document}
\vspace*{4cm}
\title{RECENT RESULTS ON THE $B_c$ MESON}

\author{ M. D. Corcoran \\ 
(for the CDF and D0 Collaborations) }

\address{Department of Physics and Astronomy, Rice University, Houston, 
TX, USA}

\maketitle\abstract{ The pseudoscalar $B_c$ is the lowest mass bound 
state of the $c\bar{b}$ and the charge conjugate. 
It is the last such meson to be discovered. 
 Both the CDF and D0 collaborations have reported 
recent results on the 
mass and lifetime of this elusive state. }

\section {Introduction}
The $B_c$ is the lowest mass bound state of $c\bar{b}$ or $b\bar{c}$.  It is 
expected to be a pseudoscalar meson, the last such meson predicted by the 
Standard Model. 
 The $B_c$  is unique in that either one of its quarks can decay, leaving
the other as a spectator. One decay path is $\bar{b} \rightarrow \bar{c} W^+$,
often leading to  final states containing a $J/\psi$. Although these modes
do not have the largest branching ratios, they have proven to be the
easiest to observe due to the clean $J/\psi$ tag. If the $c$ quark 
decays, leaving the $b$ as a spectator, the final states often 
contain $B_s$ mesons. These decay modes have yet to be observed.

There are many predictions for the mass and lifetime of the $B_c$.
Heavy quark effective potentials predict the mass to be between
6.2 and 6.3 GeV/c$^2$. \cite{quigg} Since 
the $B_c$ lies between the well-measured $b\bar{b}$ and the $c\bar{c}$ 
states, it
would be surprising if it's mass were very different from the predictions.  
Expectations for the $B_c$ lifetime are more variable, with 
predictions ranging from 0.4 to 1.4 ps, so that a measurement of the 
$B_c$ lifetime will discriminate between the different models. 
The $B_c$ is expected to have a rich spectroscopy of narrow 
excited states, which, if observable, would also constrain the heavy 
quark potential. 

  The first observation of $B_c$ came from CDF in Run I. They observed 
$B_c$ in the semileptonic decay modes $B_c \rightarrow J/\psi \mu \nu$ and 
$B_c \rightarrow J/\psi e \nu$. This result has been published for
some time.\cite{cdf1}  There are two new results from Run II, 
which are reported here. D0 has 
made an observation of the $B_c$ in the semileptonic mode 
$B_c \rightarrow J/\psi \mu X$ \cite{d01}, and CDF has evidence for the
exclusive final state $B_c \rightarrow J/\psi \pi$ \cite{cdf2}.
Both of these results are preliminary. 

\section {D0 Observation of $B_c \rightarrow J/\psi \mu X$}

The D0 result is based on 260 pb$^{-1}$ of data. The analysis proceeds 
by reconstructing a good $J/\psi$ from a dimuon data sample. A good
3-D vertex is required, and a $J/\psi$ mass constraint is applied. Then
events are selected in which  
a third muon track can be associated with the $J/\psi$ vertex. A 
background 
control sample is defined in the same way, except that the third track 
is required to not be tagged as a muon. 
  Since there is a missing neutrino in the final state, it is not 
possible to calculate the true proper time. A ``pseudo-proper time'' is
calculated event-by-event, and this distribution is corrected on average
based on Monte Carlo simulations.

  Backgrounds arise from prompt $J/\psi$s which accidently vertex with a 
third muon, and also from real heavy flavor decays which pick up a third
muon. These background samples are treated separately in the analysis, 
since they will not have the same distributions in mass or proper time.
The prompt background sample is obtained by taking events with negative
pseuo-proper time (which arise from resolution effects) and 
reflecting the distribution 
about $t=0$. This sample is then subtracted from the 
full background sample to obtain the heavy flavor background sample. 

\begin{figure} [ht]
\psfig{figure=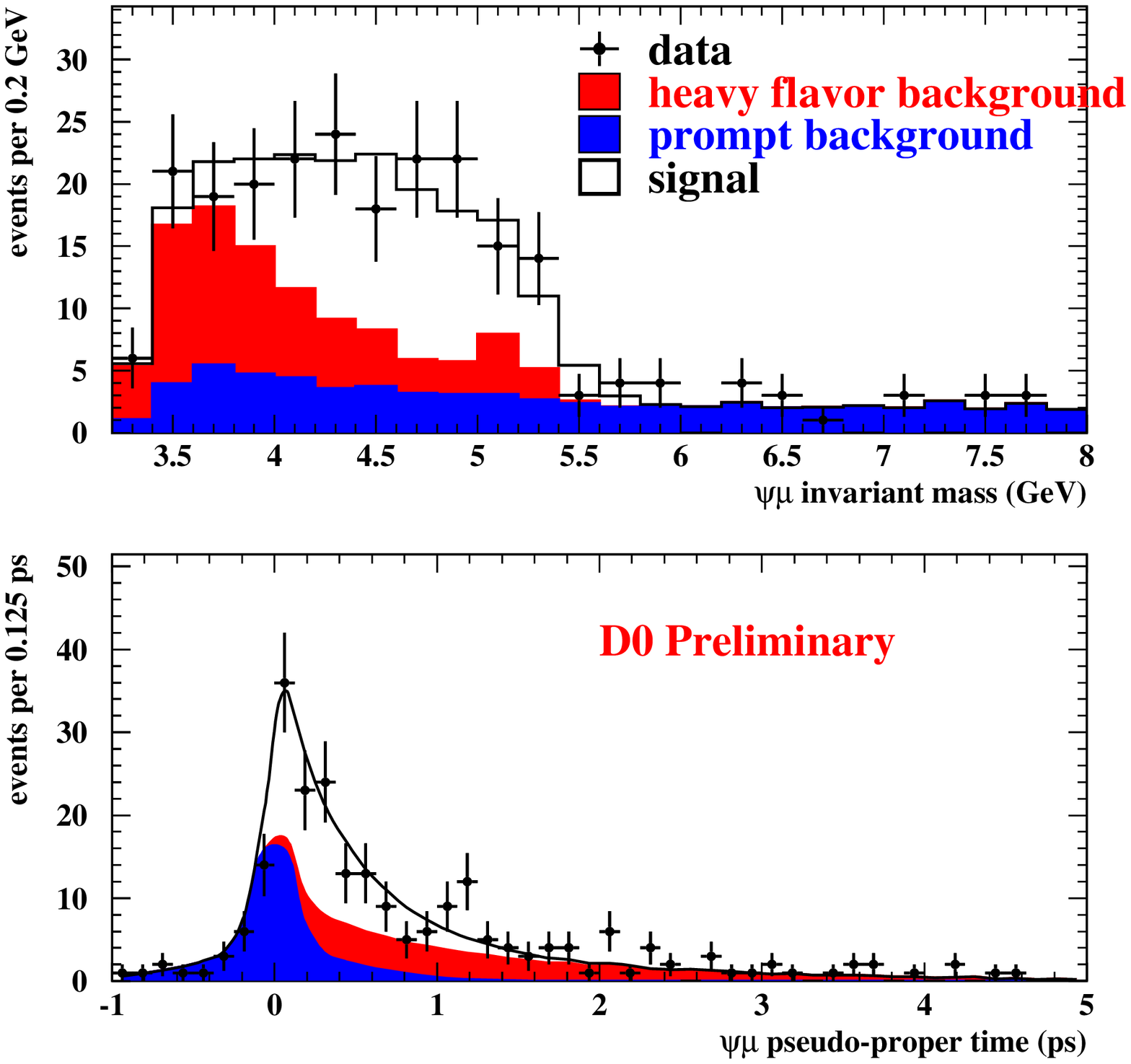,height=3.in}
\psfig{figure=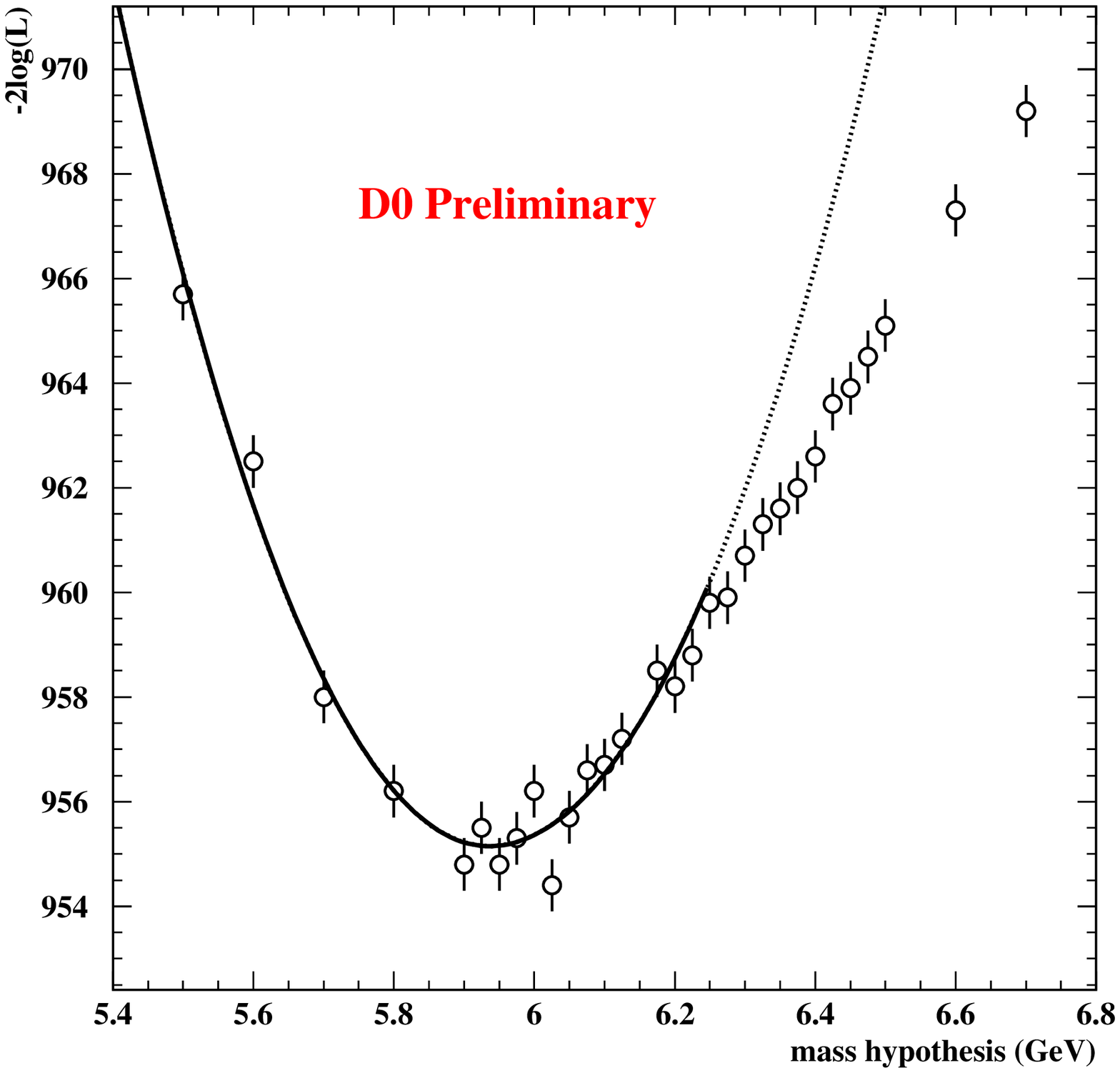,height=3in} 
\vspace*{.1in}
\caption{Results of the likelihood fit for $M_{\psi \mu}$ 
and pseudoproper time. The left plot shows the data and the fit results for
the expected backgrounds and signal contributions. The right plot is an
example log likelihood function vs. the $B_c$ mass hypothesis. }
\end{figure}

 Figure 1 shows the result of a combined 
likelihood fit for the $B_c$ mass and lifetime. The left plot shows the
distributions in 
$M_{\psi \mu}$, the invariant mass of the 
$J/\psi \mu$ system, and the pseudo-proper time. Three contributions to
the full data sample are shown: 
 the prompt background, heavy flavor background, and $B_c$ signal. 
The right plot shows an example of the log likelihood function vs.
the $B_c$ mass hypothesis, for an assumed lifetime value of
0.45 ps. The  fitting procedure gives 95 $\pm 12(stat) 
\pm 11(sys)$ signal events. 
This result is currently the most statistically significant observation of 
the $B_c$. The best fit value for the mass is $5.95 ^{+0.14}_{-0.13}(stat) 
\pm 0.34(sys)$. The best fit value for the lifetime is 
$0.45^{+0.12}_{-0.10}(stat) \pm 0.12(sys)$. The mass and lifetime are
uncorrelated in the fit.   

\section {CDF Evidence for $B_c \rightarrow J/\psi \pi$} 
 CDF has recently observed evidence for the $B_c$ in the 
exclusive final state  $B_c \rightarrow J/\psi \pi$. The analysis uses a 
360 pb$^{-1}$ sample of data with $J/\psi \rightarrow \mu \mu$ 
identified at the L3 trigger.
A mass constraint is applied to the muons forming the $J/\psi$, and a 
third track, assumed to be a $\pi$, is required to form a good 3D vertex
with the $J/\psi$. The prominent decay $B^+ \rightarrow J/\psi K^+$ is 
used as the reference mode to validate the Monte Carlo simulations 
and check the various cuts.  
The left plot in figure 2 shows the  $B^+ 
\rightarrow J/\psi 
K^+$ mass distribution, demonstrating the low level of background in 
this reference mode. The right plot of figure 2 
 shows the $M_{J/\psi \pi}$ mass distribution in the mass region 
used for the $B_c$ search.
 
\begin{figure} [ht]
\psfig{figure=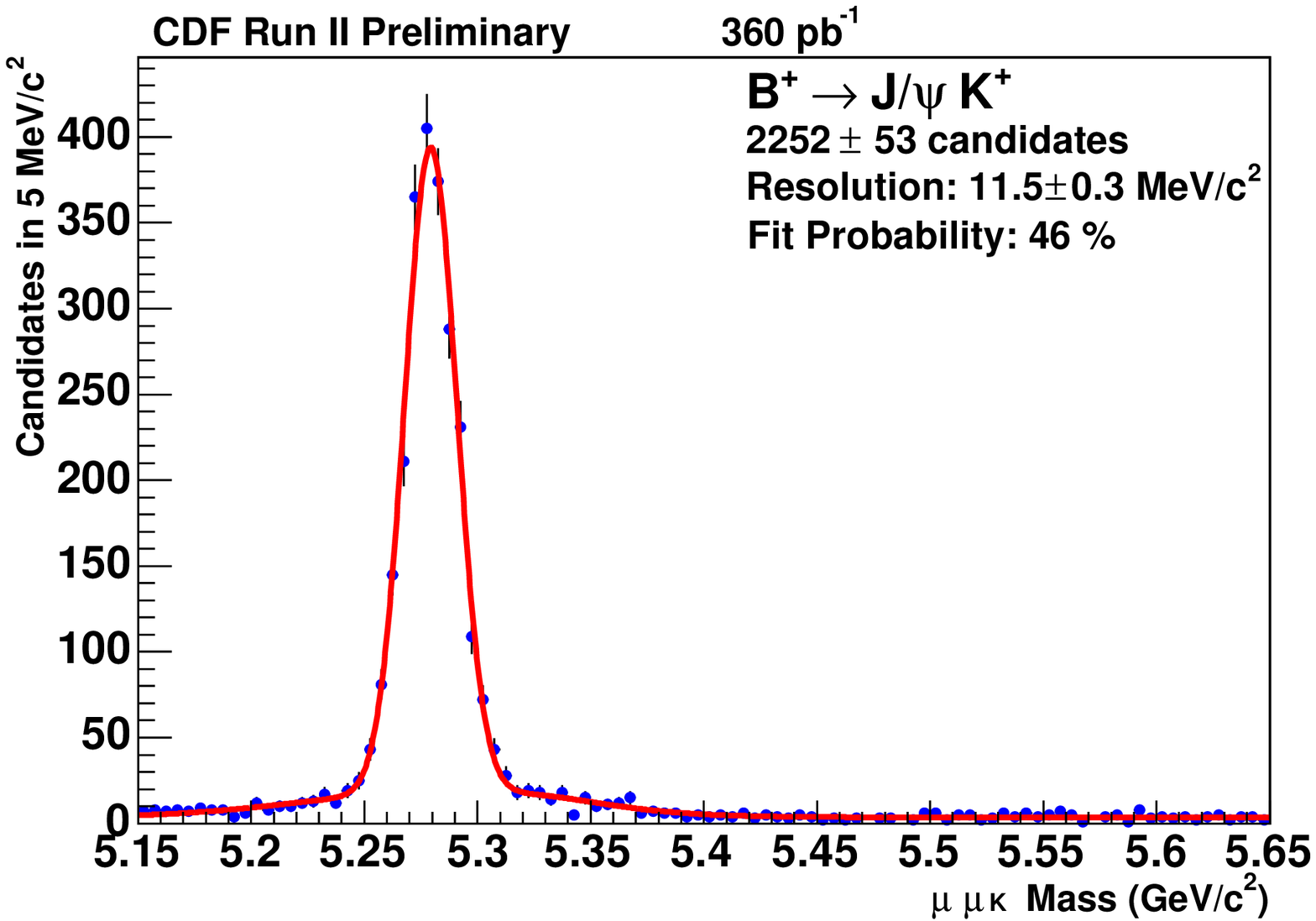,height=2.3 in}
\psfig{figure=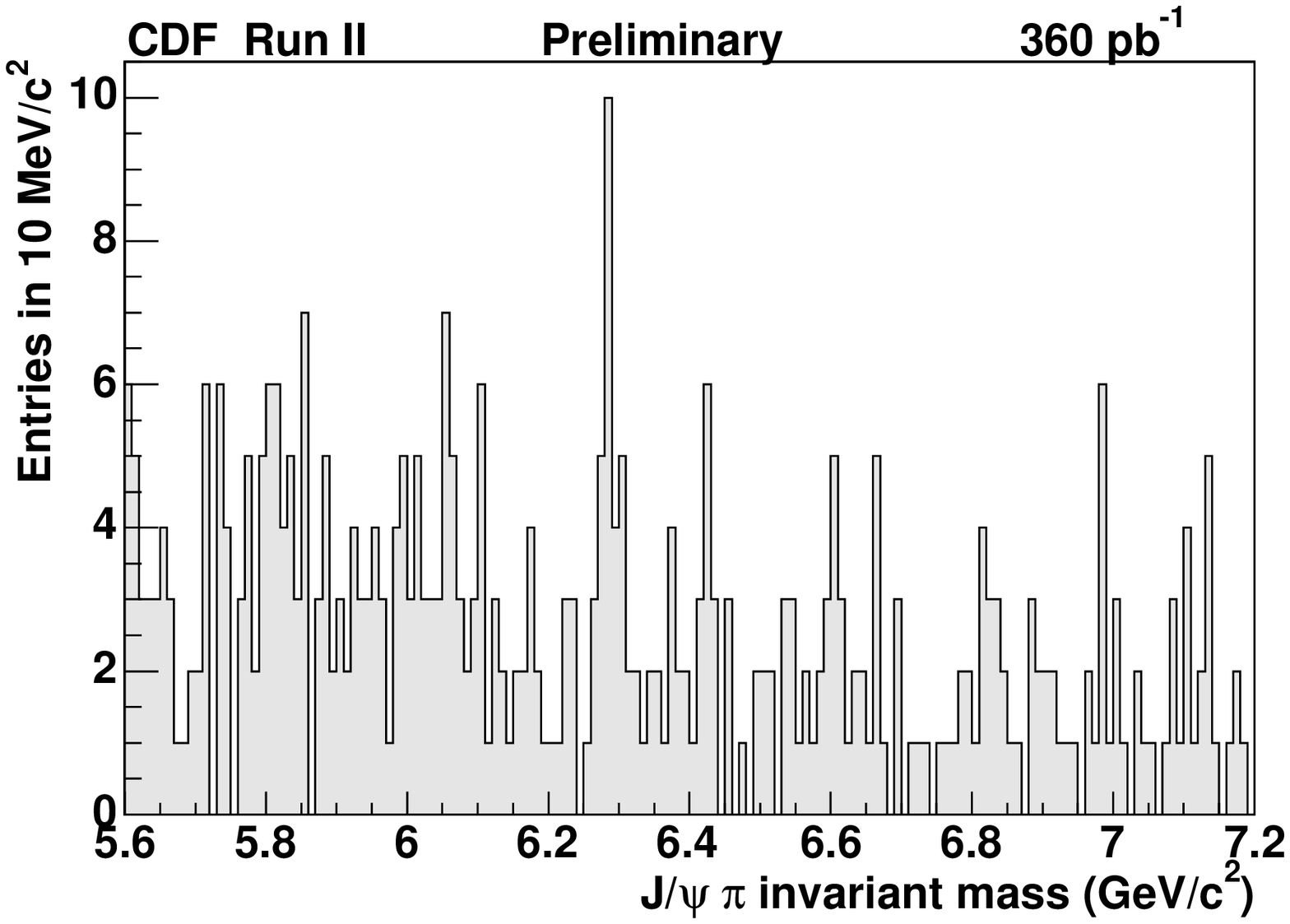,height=2.3 in} 
\vspace*{.1in}
\caption{The left plot shows the $M_{J/\psi K^+}$ distribution from 
 $B^+ \rightarrow J/\psi K^+$ reference mode. 
The right plot shows the  $M_{J/\psi \pi}$ distribution in the search region.}
\end{figure}

  A blind analysis was performed in the following way. A significance function
was defined as $\Sigma=\frac{S}{1.5+\sqrt{B}}$ where $S$ represents the
number of signal events within $\pm 2\sigma$ of the assumed mass, with 
$\sigma$ determined from Monte Carlo. $B$ is the number of background events
in the same mass region, with the background taken to be linear. 
In searching for the $B_c$ signal, a likelihood fit 
was done in bins of $M_{J/\psi \pi}$ 10 MeV/c$^2$ wide. 
 
Before the  $M_{J/\psi \pi}$ distribution in the data was revealed, 
a Monte Carlo study was performed using 1000 Monte Carlo samples which
consisted of only background. In these samples, the background consisted of
two components: a combinatorical part, which was taken to be linear, and a 
contribution from various partically reconstructed $B_c$ decays, 
taken from Monte Carlo. 
These samples were then fit for the $B_c$ signal plus background, and the 
significance
function $\Sigma$ was calculated in all mass bins from 5.7 to 7.2 GeV/c$^2$. 
 Occasionally, due to statistical fluctuations, the value of 
$\Sigma$ could be large. In these 1000 samples, over all 
mass values, the maximum
value of $\Sigma$ that occurred was 3.5. Therefore, this value of $\Sigma$
was taken as the minimum required to claim statistically 
signficant evidence for $B_c \rightarrow J/\psi \pi$. 

This entire procedure was carried out before the $M_{J/\psi \pi}$
distribution in the data was revealed. After the requirement on $\Sigma$ had
been fixed, the same procedure was carried out on the actual data. 
Figure 3 shows the value of the significance function $\Sigma$ in each 
mass bin for the data(left plot). The maximum value of $\Sigma$ observed is 
3.6, just
above the predetermined cutoff of 3.5, at a mass value around 6.3 GeV/c$^2$. 
Figure 3 also shows an
expanded view of the $M_{J/\psi \pi}$ distribution, showing
the signal region and the fit to the $B_c$ mass. The fit returns 18.9
$\pm 5.7$ signal events and a mass of 6.2879 $\pm 0.0048 (stat) 
\pm 0.0011(sys)$ GeV/c$^2$. There is as 
yet no lifetime determination from this analysis. 

\begin{figure} [ht]
\psfig{figure=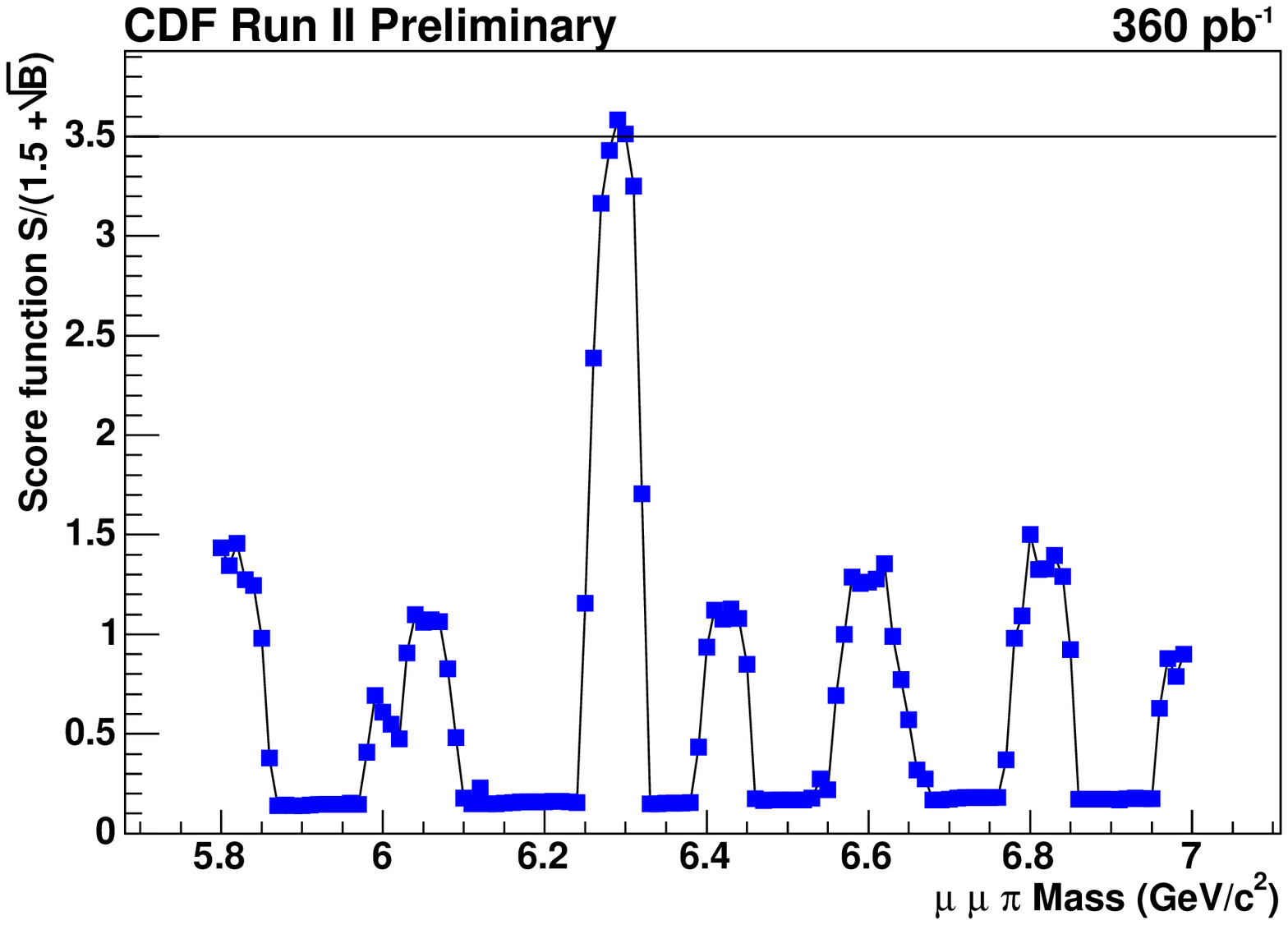,height=2.3 in}
 \psfig{figure=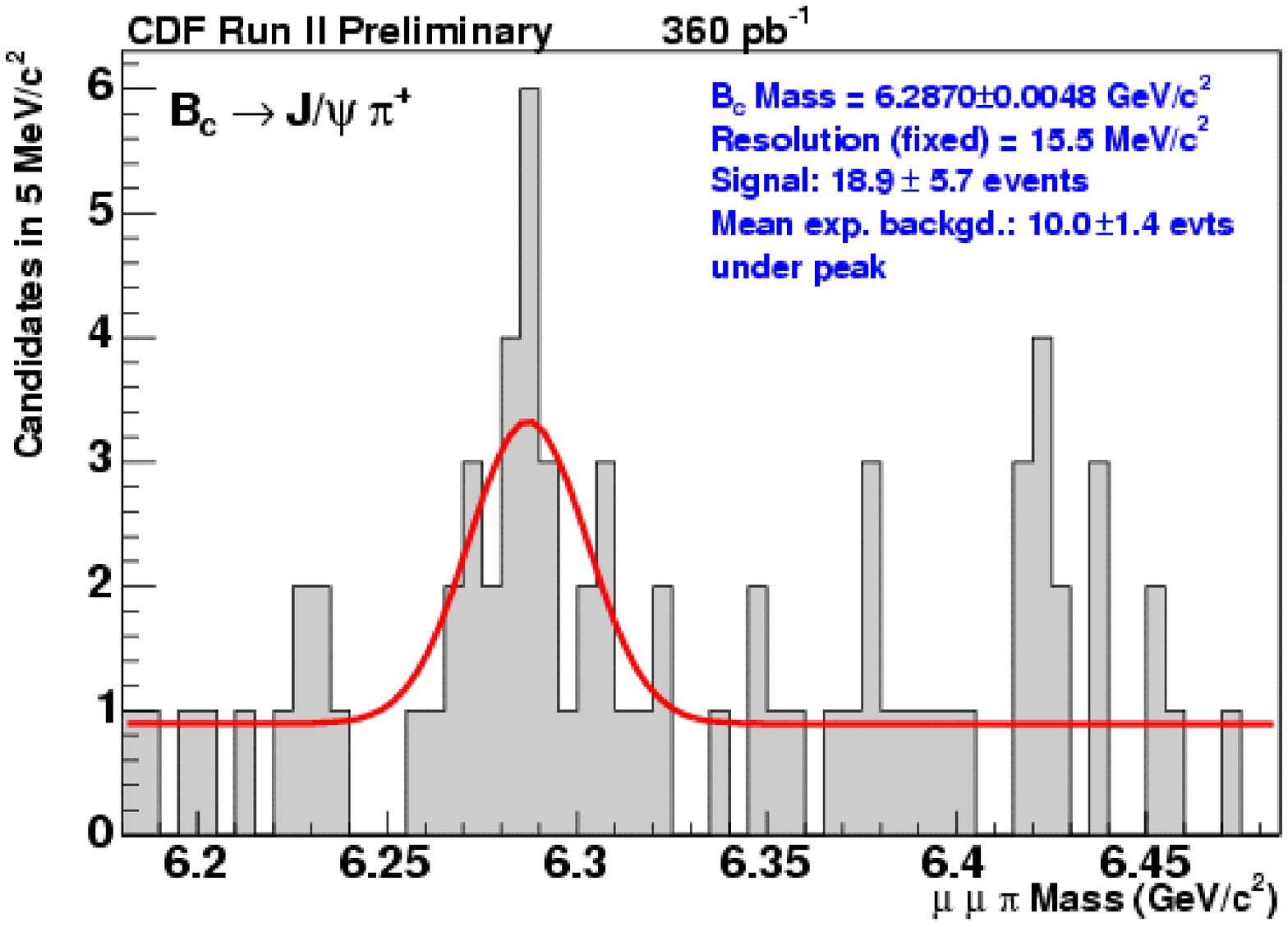,height=2.3 in}
\vspace*{.1in}
\caption{ Significance function $\Sigma$ calculated in mass bins for
the data. The maximum value of $\Sigma$ is 3.6, just above
the predetermined cutoff of 3.5. The right plot shows an expanded 
view of the $M_{J/\psi \pi}$ distribution, showing the signal
region and the result of an unbinned likelihood fit. The fit returns 18.9
$\pm 5.7$ events and a mass of 6.2879 $\pm$ 0.0048 GeV/c$^2$. }
\end{figure}

\section {Summary}

 Table I  summarizes the observations of and evidence for the
$B_c$ meson. All results are in agreement with expectations from heavy quark 
potential models. Analysis is proceeding for both experiments. 
Much more data is already in hand and we can expect improved determinations
of the $B_c$ mass and lifetime soon.

   I am grateful to Sherry Towers for
providing me with plots and information for the D0 analyisis, and 
also to Vaia Papadimitriou who was my contact for the CDF analysis. 

\begin{center}
 \begin{table}
\begin{tabular}{|c|c|c|c|c|} \hline
& & & & \\
   &  $\cal{L}$ (pb$^{-1})$ & Signal Events & Mass (GeV/c$^2$) & 
     Lifetime (ps) \\ 
& & & & \\ \hline
& & & & \\
CDF Run I & 110 & $20.4^{+6.2}_{-5.5}$  & 6.4 $\pm 0.39 \pm 0.13$ 
   & $0.46^{+0.18}_{-0.16}\pm 0.03$ \\
($B_c \rightarrow J/\psi l \nu$) & & & (published) & (published)\\ 
 & & & & \\ \hline
 & & & &   \\
D0 Run II  &  210  & $95 \pm 12 \pm 11$ & $5.95^{+0.14}_{-0.13}\pm 0.34$ 
   & $0.45^{+0.12}_{-0.10} \pm 0.12 $ \\ 
($B_c \rightarrow J/\psi \mu X$) & & & (preliminary) & (preliminary) \\
 & & & &  \\ \hline 
& &  & & \\
CDF Run II  & 360 & $18.9\pm 5.7$ & $6.287 \pm 0.0048 \pm 0.0011$ & --- \\
($B_c \rightarrow J/\psi \pi$) & & &  (preliminary)  &\\
& &  & &\\ \hline
\end{tabular}
\caption{ Summary of experimental results for the $B_c$ meson mass and 
lifetime. In all cases the first errors are statistical and the second 
are systematic.}
\end {table} 
\end{center}

\end{document}